# Machine Learning Approaches for Defect Detection in a Microwell-based Medical Device

**Xueying Zhao, Yan Chen, Yuefu Jiang, Amie Radenbaugh, Jamie Moskwa, Devon Jensen**

BD Biosciences, 155 McCarthy Blvd, Milpitas, CA

**Correspondence**

Xueying Zhao, BD Biosciences, 155 McCarthy Blvd, Milpitas, CA, 95035

Email: xueying.zhao@bd.com

**Abstract**

Microfluidic devices offer numerous advantages in medical applications, including the capture of single cells in microwell-based platforms for genomic analysis. As the cost of sequencing decreases, the demand for high-throughput single-cell analysis devices increases, leading to more microwells in a single device. However, their small size and large quantity increase the quality control (QC) effort. Currently, QC steps are still performed manually in some devices, requiring intensive training and time and causing inconsistency between different operators. A way to overcome this issue is to through automated defect detection. Computer vision can quickly analyze a large number of images in a short time and can be applied in defect detection. Automated defect detection can replace manual inspection, potentially decreasing variations in QC results. We report a machine learning (ML) algorithm that applies a convolution neural network (CNN) model with 9 layers and 64 units, incorporating dropouts and regularizations. This algorithm can analyze a large number of microwells produced by injection molding, significantly increasing the number of images analyzed compared to manual operator, improving QC, and ensuring the delivery of high-quality products to customers.

# 1. Introduction

Microfluidic devices have a wide range of applications, including in life sciences (Streets & Huang, 2013; Ortseifen *et al*., 2020) and biomechanics (Wioland *et al*, 2020; Su *et al*, 2021). In life sciences, microfluidic devices can be used for cell separation. The geometry of the microwell plays a critical role in this process as it ensures that selected cells land in specific microwells. For instance, in single-cell capture, microwells are designed to accommodate only one cell and one bead per well. Once the cell is lysed, mRNA molecules can be captured on the bead for each individual cell. With further library preparation, sequencing, and bioinformatics analysis, genomic information is obtained from each independent cell.

The recent reduction in sequencing costs has made high-throughput single-cell sequencing feasible (Yamawaki *et al*., 2021; Nguyen *et al*., 2023), which has increased the demand for upstream high-throughput cell capture. This requires more microwells in each device to capture a greater number of cells, thereby posing challenges to device quality control (QC), particularly in assessing defect rates, which is often done manually. Ensuring product quality is crucial, and visual inspections are typically conducted to prevent defects. However, manual inspection can be time consuming and requires extensive training for inspectors to accurately distinguish between defective and non-defective products. Additionally, discrepancies or inconsistencies between different inspectors can arise, further complicating the manual inspection process (Villalba-Diez *et al*., 2019).

To address these challenges, automated inspection provides a viable alternative. It can enhance inspection speed, reduce human error, and improve consistency in quality control processes (Md *et al*., 2022). Computer vision has become a key strategy for processing large volumes of images and has been applied across various industries (Feng *et al*., 2019; Paneru *et al*., 2021; Tian *et al*., 2020). Machine learning, in particular, is effective for processing large datasets and classifying them into different categories. Studies by Jung *et al*. (2021) and Khan *et al*. (2021) have demonstrated the potential of machine learning in manufacturing, including the detection of defects in 3D-printed and injection-molded parts. The application of machine learning for defect detection is also promising in other manufacturing sectors, such as semiconductors (Kim *et al*., 2012; Patel *et al*., 2020), pharmaceuticals (Ma *et al*., 2020; Raab *et al*., 2022) and life sciences (Chandra *et al*., 2023; Ghosh *et al*, 2022). Machine learning in defect detection has also been used in the industry; for example, Instrumental AI, a startup that specializes in machine learning for manufacturing inspection, has collaborated with companies like Motorola Mobility to improve the efficiency and accuracy of defect detection.

Various models have been employed to solve image classification problems, including logistic regression, support vector machine (SVM) method and convolution neural networks (CNN). Nazish *et al*. (2021) used SVM and logistic regression to classify the chest X-ray images, distinguishing between COVID-19 and normal patients. Zhang and Liu (2014) applied SVM for surface defect detection in parts. While SVM is effective for small datasets, studies have shown that CNNs are better suited for larger datasets (Liqun *et al*., 2020. For instance, Westphal and Seitz (2021) used a CNN model to classify powder bed defects in the selective laser sintering process, achieving an accuracy of 0.958 with the VGG16 model architecture.

In this work, we applied machine learning techniques to develop an automated quality control solution. Image scans of the microwells were taken, preprocessed, and used to train a convolutional neural network. Model hyperparameters were optimized using grid search and cross-validation was performed. We found that the CNN model achieved an accuracy of 0.9, indicating its effectiveness in classifying defect images versus non-defect images.

**2. Materials and Methods**.

**2.1 Dataset collection**

The device used in this paper is the BD Rhapsody™ 8-lane cartridge. Figure 1(a) presents a diagram of the cartridge, which contains eight lanes for input of eight samples. In each lane, there are hundreds of thousands of microwells to capture individual cells. Figure 1(b) shows an image of a non-defect well (labeled as 0), while Figure 1(c) and 1(d) displays examples of defect wells, both labeled as 1. These images were captured using the BD Rhapsody™ scanner, which is equipped with a Basler aca3800-14µm camera. The camera has a minimum resolution is 3840 x 2748 pixels, a default frame rate of 14 fps, and an exposure time of 100 milliseconds. Due to the large field of view in scanner, the images were preprocessed into smaller 111 x 111 pixels images prior to input into the model.

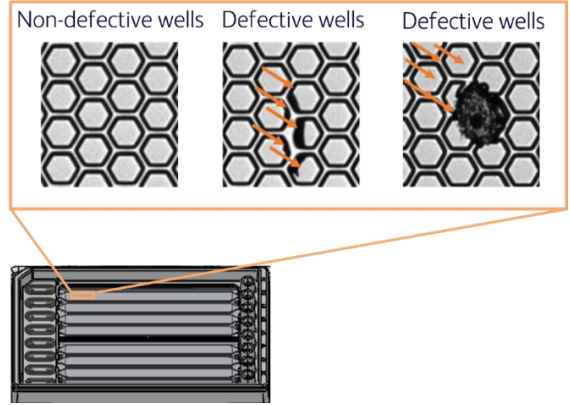

**Figure 1**. Diagram of the BD Rhapsody™ 8-lane cartridge and comparison between non-defective wells vs defective microwells

**2.2 Model structure**

    Microwell images were categorized into two classes: non-defective wells labeled as OK or 0 and defective wells labeled as NG or 1. Data augmentation techniques, such as mirroring and flipping, were applied, increasing the total dataset size to 1,000 images (500 defective microwell images and 500 non-defective microwell images). These data were used to train a logistic regression model and a neural network model. An overview of the neural network model training is presented in Figure 2 (Khan *et al.*, 2021), with the detailed architecture of the model used in this study shown in Figure 3. Hyperparameter tuning was conducted using grid search, and selected parameter values are listed in Table 1. The CNN model consists of 9 layers with 48 units and a train/validation split of 0.2. Cross validation was also performed to enhance the model's generalization. The trained model was subsequently used to predict whether new input images represented non-defective or defective wells.

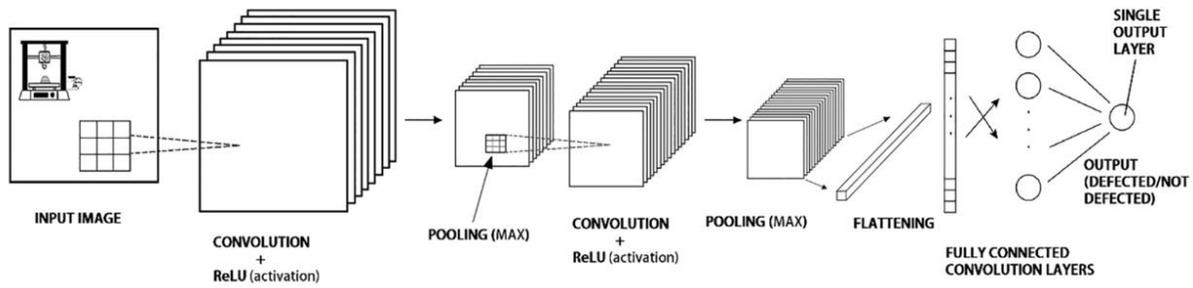

**Figure 2**. Training Model of Convolutional Neural Network (Khan *et al*., 2021)

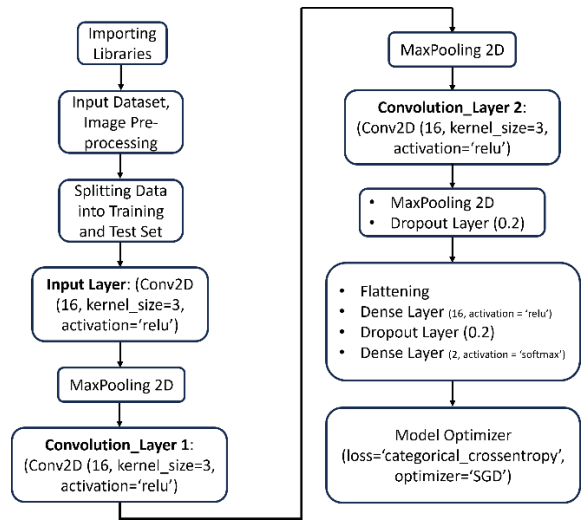

**Figure 3**. Structure of the Convolutional Neural Network Model used in this study

Table 1. Hyperparameters used in this study

| Cost function | Learning rate | Optimizer | No. Epochs | Batch size | Dropout rate | L2 regularization |
|---|---|---|---|---|---|---|
| Sparse Categorical Cross Entropy | 0.001 | Adam | 40 | 16 | 0.2 | 0.3 |

## 2.3 Performance measures

The performance of the two models was compared using the following metrics: accuracy, precision, recall, and F1 score. These metrics were calculated based on the values of true positive (TP: number of images correctly classified as positive), true negative (TN: number of images correctly classified as negative), false positive (FP: number of images incorrectly classified as positive), and false negative (FN: number of images incorrectly classified as negative). The definitions for accuracy, precision, recall, and F1 score are as follows:

$$\text{Accuracy} = \frac{\#\ correctly\ classified\ images}{\#\ all\ images} = \frac{TP+TN}{TP+FP+TN+FN}$$

$$\text{Precision} = \frac{\#\ images\ correctly\ classified\ as\ positive}{\#\ images\ classified\ as\ positive} = \frac{TP}{TP+FP}$$

$$\text{Recall} = \frac{\#\ images\ correctly\ classified\ as\ positive}{\#\ images\ classified\ as\ negative} = \frac{TP}{TP+FN}$$

$$\text{F1 score} = 2 \times \frac{precision \times recall}{precision+recall} = \frac{2 \times TP}{2 \times TP+FP+FN}$$

## 3. Results and discussion

As described in the performance measures section, accuracy, precision, recall and F1 score were used to evaluate the models. Table 2 shows the comparison between the logistic regression and CNN models. As expected, the logistic regression model reported lower accuracy compared to the CNN model, as logistic regression typically performs better with linear relationships. Given the increased complexity of image data, a more sophisticated model, such as a CNN, is necessary for better performance.

**Table 2**. Performance comparison between logistic and CNN models

| Method | Accuracy | Precision | Recall | F1 score |
|---|---|---|---|---|
| Logistic Regression | 0.79 | 0.81 | 0.75 | 0.78 |
| CNN | 0.90 | 0.92 | 0.88 | 0.90 |

As shown in Figure 4, the CNN model employed early stopping at 25 epochs. The training and validation accuracy, as well as loss against epochs, are depicted in Figure 4. As the number of epochs increases, the accuracy improves and stabilizes around 0.9 for the CNN model. Simultaneously, the validation loss curve decreases from 1.0 to approximately 0.3, indicating that the model is able to generalize well on new data.

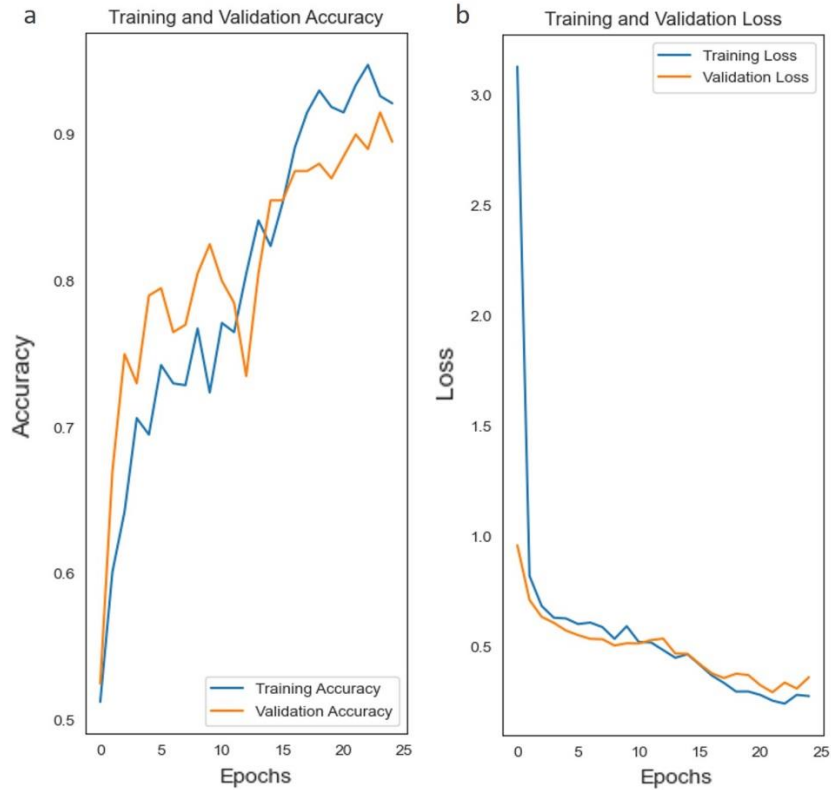

**Figure 4**. Training and validation accuracy and loss of the CNN model. (a) accuracy vs epochs. (b) loss vs epochs.

## 4. Conclusions and future research

In this paper, both the logistic regression model and CNN model were employed to classify defective vs non-defect wells as part of a quality check for a microfluidic device. Hyperparameter tuning was applied to the CNN model to enhance its performance. The CNN model reported a higher accuracy of up to 0.9 compared to the logistic regression model. This approach allows the screening of more samples than manual counting while maintaining consistent results, demonstrating the value of integrating machine learning models into quality control processes to enhance efficiency and consistency. Future work will involve multi-class classification to identify various types of defects, which will provide feedback to manufacturing processes and potentially inform design improvements to enhance the quality of the device.

Md, A. Q., Jha, K., Haneef, S., Sivaraman, A. K., & Tee, K. F. (2022). A review on data-driven quality prediction in the production process with machine learning for industry 4.0. *Processes*, *10*(10), 1966.

Nazish, Ullah, S. I., Salam, A., Ullah, W., & Imad, M. (2021, March). COVID-19 lung image classification based on logistic regression and support vector machine. In *European, Asian, Middle Eastern, North African Conference on Management & Information Systems* (pp. 13-23). Cham: Springer International Publishing.

Nguyen, R., Hlathu, Z., Gordon, C., Zhao, X., Moskwa, J., Jensen, D., ... & Ayer, A. (2023). Flexible and high-throughput approach for capturing large number of single cells simultaneously using microwell-based technology. *The Journal of Immunology*, *210*(1_Supplement), 251-08.

Ortseifen, V., Viefhues, M., Wobbe, L., & Grünberger, A. (2020). Microfluidics for biotechnology: bridging gaps to foster microfluidic applications. *Frontiers in Bioengineering and Biotechnology*, *8*, 589074.

Paneru, S., & Jeelani, I. (2021). Computer vision applications in construction: Current state, opportunities & challenges. *Automation in Construction*, *132*, 103940.

Patel, D. V., Bonam, R., & Oberai, A. A. (2020). Deep learning-based detection, classification, and localization of defects in semiconductor processes. *Journal of Micro/nanolithography, MEMS, and MOEMS*, *19*(2), 024801-024801.

Raab, D., Fezer, E., Breitenbach, J., Baumgartl, H., Sauter, D., & Buettner, R. (2022, June). A Deep Learning-Based Model for Automated Quality Control in the Pharmaceutical Industry. In *2022 IEEE 46th Annual Computers, Software, and Applications Conference (COMPSAC)* (pp. 266-271). IEEE.

Tian, H., Wang, T., Liu, Y., Qiao, X., & Li, Y. (2020). Computer vision technology in agricultural automation—A review. *Information Processing in Agriculture*, *7*(1), 1-19.

Streets, A. M., & Huang, Y. (2013). Chip in a lab: Microfluidics for next generation life science research. *Biomicrofluidics*, *7*(1).
10